# GPU-accelerated single-cell analysis at scale with rapids-singlecell

Severin Dicks[1,2*], Lukas Heumos[1,3], Lilly May[1,4], Sara Jimenez[1], Philipp Angerer[1], Ilan Gold[1], Isaac Virshup[1], Felix Fischer[1,5], Michelle Gill[2], Melanie Boerries[6,7], Corey J Nolet[2,8], Tiffany J. Chen[2], Fabian J. Theis[1,3,5*]

[1]Institute of Computational Biology, Helmholtz Center Munich, Germany
[2]NVIDIA Corporation
[3]TUM School of Life Sciences Weihenstephan, Technical University of Munich, Germany
[4]Department of Experimental Psychology, University of Oxford, United Kingdom
[5]TUM School of Computing, Information and Technology, Technical University of Munich, Munich, Germany
[6]Institute of Medical Bioinformatics and Systems Medicine, Medical Center - University of Freiburg, Faculty of Medicine, University of Freiburg, Freiburg, Germany
[7]German Cancer Consortium (DKTK), Partner site Freiburg, a partnership between DKFZ and Medical Center - University of Freiburg, Germany
[8]Department of Computer Science & Electrical Engineering, University of Maryland Baltimore County
[*]Correspondence to sdicks@nvidia.com and fabian.theis@helmholtz-muenchen.de

Single-cell sequencing technologies reveal cellular heterogeneity at high resolution, advancing our understanding of biological complexity. As datasets start to scale to tens of millions of cells, computational workflows face substantial bottlenecks, with CPU-based analytical pipelines requiring hours or days for routine processing steps like filtering, normalization, and clustering. These scalability limitations fundamentally restrict common interactive data exploration and iterative hypothesis testing. Here we introduce rapids-singlecell, a GPU-accelerated framework that integrates natively with the scverse ecosystem and operates directly on the AnnData data structure, which delivers orders-of-magnitude speedups for single-cell workflows. Built on CuPy arrays and the NVIDIA CUDA-X Data Science (RAPIDS) ecosystem, rapids-singlecell provides near drop-in GPU replacements for core scanpy-based analysis steps. Across standard single-cell workflows such as preprocessing, dimensionality reduction, neighborhood graph construction, clustering, and batch correction, rapids-singlecell achieves speedups of up to several hundred-fold compared to optimized CPU baselines. This reduces analysis time from hours to minutes on standard hardware, while maintaining consistent biological interpretations. These performance improvements make it possible to analyze large data sets in close to real time, without the need for data splitting. Together with real-time parameter tuning and iterative workflows, rapids-singlecell makes interactive large-scale single-cell analysis possible.

# GPU-accelerated single-cell analysis at scale with rapids-singlecell

Single-cell sequencing technologies have advanced our understanding of biological complexity by resolving cellular heterogeneity at single-cell resolution[1]. Yet, as datasets grow to tens of millions of cells, computational workflows face severe scalability challenges. CPU-based analytical pipelines create significant bottlenecks for routine steps like filtering, normalization, dimensionality reduction, and clustering, slowing iterative data exploration and parameter tuning.

Many core single-cell analysis operations - matrix transformations, nearest-neighbor searches, and clustering - are inherently parallelizable. High-performance computing approaches, particularly those leveraging graphics processing units (GPUs), can address these bottlenecks. GPUs excel at massively parallel operations, enabling orders-of-magnitude speedups for data processing tasks such as linear algebra, graph computations, and machine learning[2]. Although GPU-accelerated libraries are widely used in deep learning, their adoption in single-cell omics pipelines has been limited, partly because established toolchains have not integrated GPU computing seamlessly into existing data structures or workflows.

Here, we introduce **rapids-singlecell**, a GPU-accelerated framework that enables interactive single-cell analysis at scale. Rapids-singlecell integrates into the scverse[3] ecosystem and operates directly on the AnnData[4] data structure, a community standard (**Figure 1a**). Built on the NVIDIA and scverse ecosystems (**Figure 1b**), rapids-singlecell accelerates single-cell workflows by up to several orders of magnitude, reducing runtimes from hours on multi-core CPUs to minutes or seconds on a single GPU, thereby transforming large-scale single-cell analysis from a delayed, batch-driven process into an interactive workflow (**Figure 1c**).

Rapids-singlecell is accessible as an open-source package hosted at https://github.com/scverse/rapids-singlecell and installable via pip or Conda. We provide comprehensive documentation at https://rapids-singlecell.readthedocs.io, including tutorials, hardware requirements, and use-cases covering workflows from basic preprocessing to multi-million-cell analyses with Dask-based multi-GPU scaling. Additional guidance covers memory management, oversubscription settings, and transitioning existing scanpy[5] scripts to GPU-powered versions through rapids-singlecell.

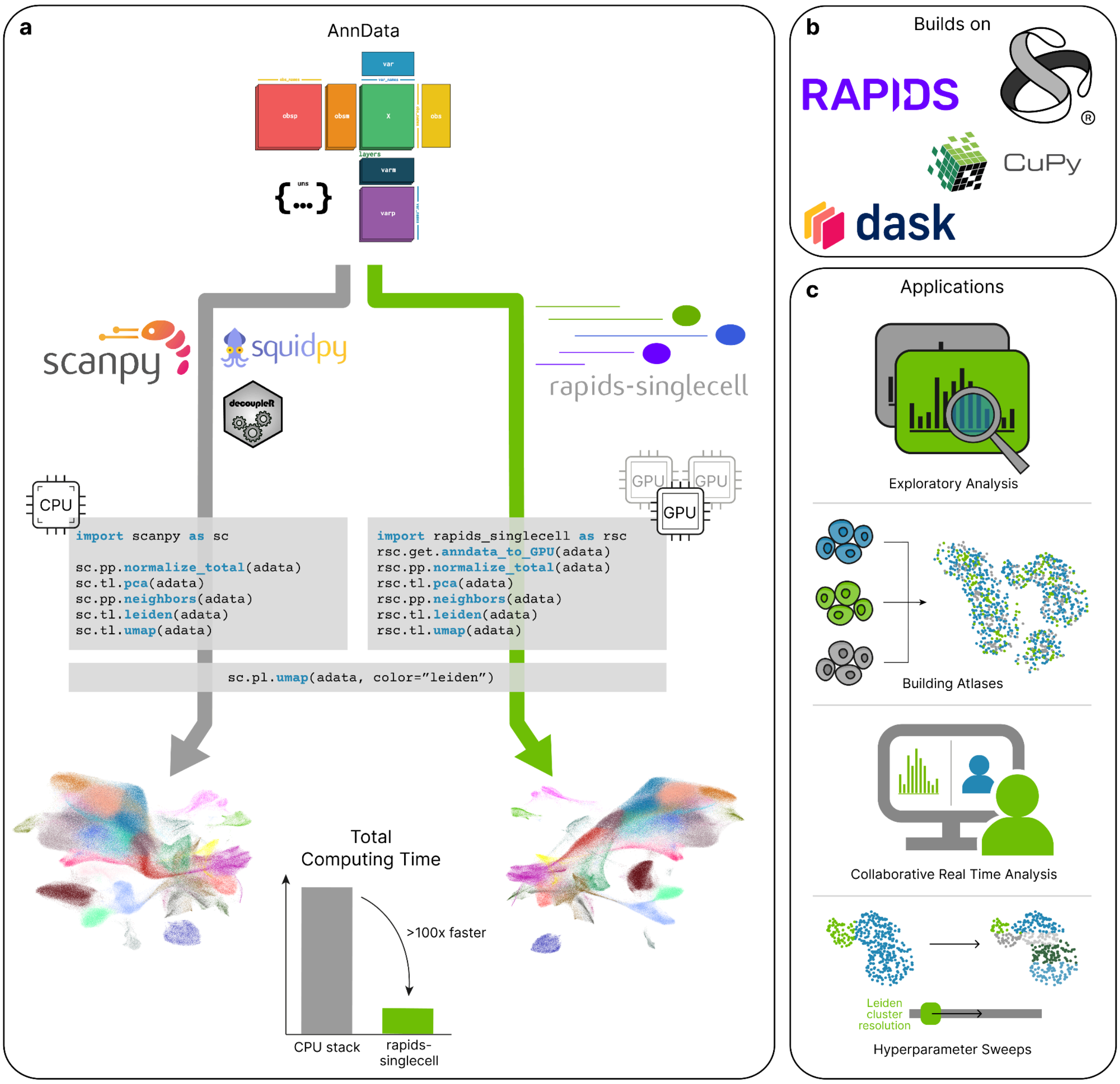

*Figure 1.* ***Overview of the rapids-singlecell workflow.*** *(a) rapids-singlecell integrates with the scverse ecosystem, sharing the AnnData data structure with existing frameworks. The API mirrors scanpy, decoupler[6], pertpy[7], and squidpy[8], requiring minimal code changes to transition from CPU to GPU execution. A single function transfers data to GPU memory, after which standard preprocessing, dimensionality reduction, and visualization steps achieve more than 100× speedup. (b) The framework builds on NVIDIA CUDA-X Data Science (RAPIDS), CuPy, and Dask for GPU-accelerated computation and out-of-core processing. (c) The speedups enable target applications including interactive exploratory analysis, massive cell atlas construction, collaborative real-time analysis, and routine parameter sweeps on million-cell datasets that are impractical with CPU-based workflows.*

# Technical architecture

At the core of rapids-singlecell is CuPy[2] (**Figure 1b**), selected over frameworks like PyTorch[9] or JAX[10] for its robust support of sparse matrices and support for the Array API, allowing users to write code similar to standard NumPy[11] and SciPy[12] sparse while executing on GPUs. Sparse data handling is critical as gene expression matrices typically contain many zeros. By operating directly on sparse representations throughout the analysis pipeline, rapids-singlecell avoids matrix densification and repeated host-device transfers that would otherwise dominate runtime at scale. CuPy's sparse array implementations, closely modeled after SciPy's data structures, deliver efficient memory usage and fast linear algebra operations on large datasets while preserving familiar NumPy and SciPy syntax. This design minimizes adoption friction for users of existing CPU-based pipelines. To enable CuPy arrays in AnnData, we collaborated with scverse to improve the adoption of the array API standard.

Built on CuPy arrays, rapids-singlecell leverages the broader RAPIDS[13] ecosystem, including cuML[14] for machine learning tasks (e.g., PCA, UMAP[15]), cuGraph for graph-based computations (e.g., clustering), and cuVS for vector search (**Methods**). To further optimize performance, we use CuPy CUDA RawKernels – just-in-time (JIT) compiled CUDA kernels – that allow writing custom GPU code directly in Python and C (**Methods**). This fine-grained control over kernel operations reduces data transfers, optimizes intermediate steps, and avoids unnecessary memory copies. Custom GPU kernels streamline operations like gene selection and sparse gram matrix computation, avoiding repeated CPU-GPU transfers and intermediate array allocations.

Memory management is also enhanced through the RAPIDS Memory Manager (RMM), which orchestrates GPU allocations and supports oversubscription. When GPU memory is limited, RMM automatically spills data to host memory, allowing analyses that exceed local GPU capacity to proceed without manual intervention. While some speed reduction is inevitable when oversubscribing memory, this approach enables analyses of hundreds of thousands of cells on consumer-grade GPUs.

For massive datasets beyond the scale of a single GPU, rapids-singlecell integrates Dask[16], a parallel computing framework that facilitates out-of-core and multi-GPU computations (**Methods**). Dask enables dynamic task scheduling and parallel execution across multiple GPUs and even multiple nodes, splitting large datasets into manageable chunks and distributing both data and computations. This out-of-core strategy bypasses CuPy's reliance on int32 for sparse CSR index arrays (indptr, indices), which currently impose a hard ceiling on non-zero entries and create practical memory bottlenecks around one million cells for typical stencil-based discretizations. By employing Dask, users can scale analyses to millions of cells and beyond, limited primarily by the aggregate memory and compute resources available.

## Performance and reproducibility

We benchmarked rapids-singlecell on 1 million cells from the 10x Genomics mouse brain dataset[17], running a standard workflow: preprocessing, normalization, HVG selection, dimensionality reduction (PCA, UMAP, t-SNE), neighborhood graph construction, and Leiden clustering (**Methods**).

On a 32-core workstation, the pipeline took over 52 minutes (**Methods**). A single NVIDIA DGX B200 GPU completed it in 26 seconds - a >120 speedup, placing end-to-end million-cell analysis within interactive timescales. Individual steps showed even larger gains: ~70× for preprocessing, ~350× for UMAP, and ~100× for Leiden clustering (**Figure 2**).

To assess performance across architectures, we also benchmarked the pipeline on NVIDIA RTX PRO 6000 (Blackwell Workstation Edition) and NVIDIA H200 GPUs. The RTX PRO 6000 (Blackwell WS) achieved a 120× speedup, while the H200 reached 110×, both relative to the CPU baseline. All steps on the H200, RTX PRO 6000 (Blackwell WS) and B200 were performed using pooled memory allocation (**Figure 2**).

We further benchmarked GPU-accelerated batch correction using the Harmony algorithm integrated in rapids-singlecell. Performance gains were consistent across architectures: we observed a roughly 100× speedup on smaller datasets (~200,000 cells) across all tested GPUs. These gains scaled with dataset size, where all architectures (B200, H200, and RTX PRO 6000 Blackwell WS) achieved a >250× speedup on a PCA embedding of 2.4 million cells relative to the CPU baseline. To test scalability, we further ran Harmony on a PCA embedding of 11.4 million cells. All tested GPUs completed the correction in under 25 seconds, whereas a CPU-based version did not complete within a two-hour time limit and was subsequently aborted (**Figure 2, Methods**).

A new Dask integration co-developed with the scanpy developers - available in both scanpy and rapids-singlecell - enables out-of-core computation through chunk-based execution, allowing analysis of datasets that exceed available GPU memory. Using Dask also facilitates multi-GPU execution by default and further extends scalability, enabling the analysis of the 100 million cell Tahoe 100M dataset[18] in less than 20 minutes (**Methods**).

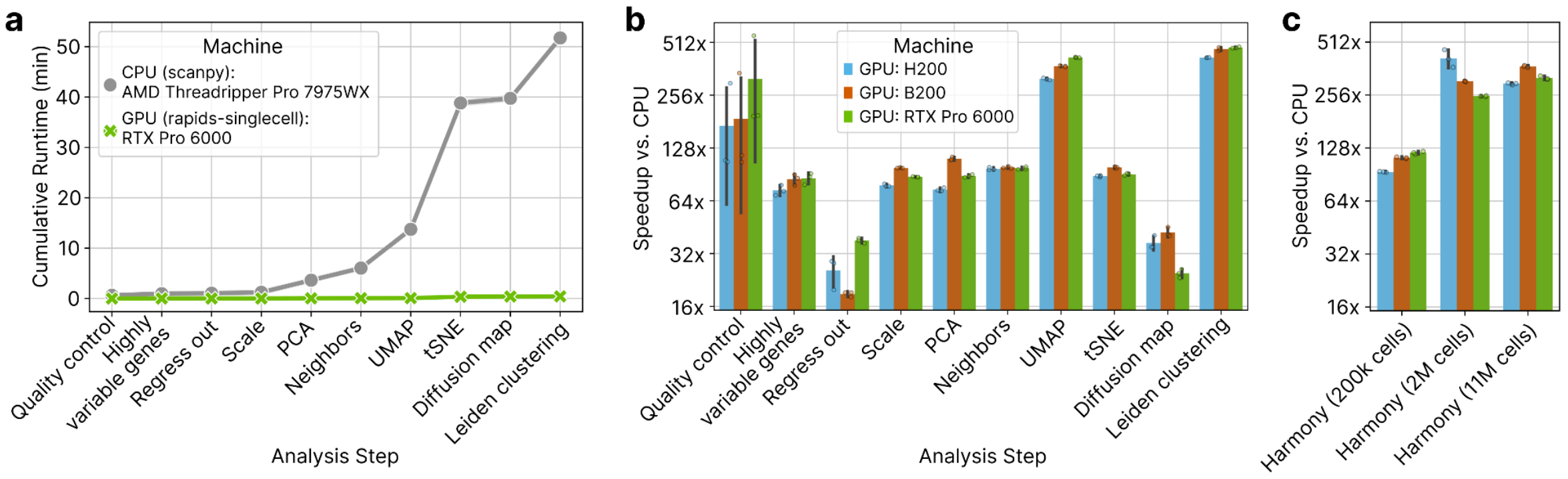

*Figure 2. **Benchmark comparison of CPU versus GPU execution across single-cell analysis steps.** (a) Cumulative runtime for a standard preprocessing and analysis workflow on a 1.3 million cell dataset. CPU execution (scanpy, AMD Threadripper Pro 7975WX) takes ~52 minutes, while GPU execution (rapids-singlecell, NVIDIA RTX PRO 6000 Blackwell workstation edition) completes in ~25 seconds. Curves represent the mean of N = 3 independent runs; standard deviation is plotted but not visible due to its small magnitude. (b) Runtime speedup of different GPU hardware (NVIDIA H200, B200, RTX PRO 6000) relative to CPU. Bars show mean speedup across N = 3 independent runs; error bars represent standard deviation. Preprocessing and analysis of 1.3 million cells, corresponding to the workflow in (a). Steps that are major bottlenecks on CPU - neighbor graph construction, clustering, and dimensionality reduction - show the largest speedups on GPU. (c) Same as (b), but for harmony batch integration for datasets of ~200,000, 2 million, and 11 million cells.*

Rapids-singlecell is designed to produce results consistent with established CPU-based workflows like scanpy[5]. Minor numerical differences arise from GPU parallelization, floating-point precision, and non-deterministic atomic operations inherent to GPU computing[19]. We validated preprocessing functions (normalization, log transformation, scaling) against CPU baselines and confirmed numerically equivalent outputs within floating-point tolerance. For stochastic algorithms (PCA, neighbor graph construction, UMAP, clustering), exact numerical reproducibility is not expected; however, downstream biological interpretations - including cluster assignments, marker gene rankings, and cell type annotations - remain consistent between implementations (**Methods**).

Despite massive benefits for large datasets, GPU acceleration provides diminishing returns for small datasets, highlighting that rapids-singlecell is specifically designed for the large-scale regime increasingly common in modern single-cell studies[20]. For fewer than approximately 50,000 cells, kernel launch overhead and host-device memory transfers can offset computational gains, making CPU execution comparable or faster. Similarly, workflows dominated by many small sequential operations benefit less from GPU parallelism. On memory-constrained consumer GPUs (<8 GB VRAM), automatic oversubscription to host memory preserves functionality but substantially reduces performance gains. Finally, operations not yet implemented in rapids-singlecell require fallback to CPU execution, incurring data transfer overhead; users should profile mixed workflows to assess overall benefit. To make GPU acceleration automatically accessible, we're further planning to automatically delegate function calls in other scverse frameworks like scanpy to rapids-singlecell where possible.

Hardware selection for *rapids-singlecell* is driven by the computational requirements of high-dimensional, sparse single-cell data. To achieve significant speedups, our framework relies on the optimized sparse matrix primitives, graph algorithms, and additional GPU-accelerated algorithms, with CUDA-X Data Science (RAPIDS) serving as a foundation, having already been established in production environments. This approach currently relies on CUDA-capable hardware, and supports a broad range of compute options suitable for different resource needs. Consumer GPUs (e.g., RTX 5090) provide an accessible entry point; while bounded by local memory, they can process datasets scaling to millions of cells by leveraging the framework's Dask-integrated out-of-core capabilities. For heavy workloads requiring substantial on-device memory without the infrastructure demands of datacenter accelerators (e.g., B200), professional workstation cards such as the RTX PRO 6000 (Blackwell WS) offer a highly effective balance of performance and memory density.

Finally, exact numerical reproducibility between CPU and GPU implementations is not guaranteed for stochastic algorithms due to differences in floating-point accumulation order, random number generation, and non-deterministic atomic operations. While biological interpretations remain consistent, users requiring bitwise reproducibility for regulatory or audit purposes should document hardware configurations and software versions explicitly.

# Conclusion

By reducing runtimes from hours to seconds, rapids-singlecell enables interactive exploratory analysis such as parameter sweeps, threshold adjustments, and embedding comparisons that were previously impractical. For clinical and translational settings, such acceleration can significantly shorten the time between data generation to actionable results.

As single-cell atlases scale to tens of millions of cells and spatial transcriptomics datasets expand in resolution, GPU acceleration will likely play an increasingly central role. Future work will aim to extend support for spatial transcriptomics, integrating with emerging foundation models for cell biology, and improve across a wider range of hardware platforms to improve access. The modular architecture is designed to allow new GPU-accelerated methods to be incorporated as the field evolves. By maintaining API compatibility with scanpy and the broader scverse ecosystem, rapids-singlecell enables users to adopt GPU acceleration stepwise, minimizing disruption to established workflows.

# Methods

## Implementation of rapids-singlecell

Rapids-singlecell is implemented in Python and builds upon several scientific open-source libraries. GPU acceleration is provided through CuPy[2], a NumPy[11]-compatible array library with support for sparse matrices, and the CUDA-X Data Science (RAPIDS) ecosystem, including cuML[14] for machine learning algorithms (PCA, UMAP, t-SNE), cuGraph for graph computations (Leiden clustering), and cuVS for vector search (eg. neighbors). Memory management and oversubscription are handled by the RAPIDS Memory Manager (RMM). Out-of-core and multi-GPU scaling are enabled by Dask. rapids-singlecell integrates tightly with the scverse ecosystem by operating directly on the AnnData data structure.

Rapids-singlecell operates on sparse matrices in Compressed Sparse Row (CSR) and Compressed Sparse Column (CSC) formats. CSR format is used for cell-wise operations (e.g., normalization, QC metrics), where efficient row access is required, while CSC format is used for gene-wise operations (e.g., highly variable gene selection), where column access dominates. Format conversions are performed lazily and cached to minimize overhead.

All functions in rapids-singlecell follow the algorithmic implementations of scanpy, ensuring that users obtain equivalent biological results when transitioning from CPU to GPU workflows. In the following sections, we primarily describe only those components where the GPU implementation necessarily differs from the scanpy CPU version due to hardware-specific optimizations or library constraints.

## Preprocessing and normalization

Raw count matrices are stored in AnnData objects following Scanpy conventions. Library size normalization rescales cell counts to a common target sum $s$: $x'_{ij} = (x_{ij} / \Sigma_j x_{ij}) \cdot s$, where $x_{ij}$ is the raw count of gene $j$ in cell $i$. Logarithmic transformation stabilizes variance: $x''_{ij} = log(1 + x'_{ij})$. Highly variable gene selection is based on per-gene mean and variance, computed using a custom GPU reduction kernel that operates directly on sparse matrices, reducing overhead compared to sequential CuPy reductions. Some of our custom CUDA kernels for preprocessing were reverse-ported to scanpy via Numba to make those workloads multi-threaded on CPU.

## Principal component analysis

PCA is performed on sparse matrices without densification. For $X \in \mathbb{R}^{nxp}$ (cells × genes, CSR format), per-feature means are computed as $\mu = (1/n)X^\top \mathbf{1}$. The upper triangle of $X^\top X$ is accumulated using a custom kernel and symmetrized via a branch-free copy kernel. The mean-centered covariance is computed implicitly as $C = (1/(n-1))(S - n\mu\mu^\top)$, avoiding explicit centering of the sparse matrix. Eigen-decomposition yields the top-$k$ loadings $V \in \mathbb{R}^{pxk}$. The sparse-to-dense transform applies implicit centering in chunks: $Z = XV - \mathbf{1}(\mu^\top V)$. We also provide uncentered truncated SVD for applications where centering is not required. We note that the Dask sparse PCA algorithm, originally engineered here for GPU acceleration, was subsequently adapted for CPU execution and integrated into scanpy. It now serves as scanpy's default solver for large-scale sparse analysis.

## Dimensionality reduction and graph analysis

Non-linear embeddings (UMAP, t-SNE) use cuML implementations. k-nearest neighbor graph construction and community detection (Leiden clustering) are accelerated through cuGraph. Results are stored in AnnData *.obsp* slots to maintain compatibility with scanpy.

## Spatial autocorrelation metrics

Rapids-singlecell accelerates spatial statistics from Squidpy[8]. Moran's I measures global spatial autocorrelation: $I = (n/W) \cdot (\Sigma_i \Sigma_j w_{ij}(x_i - \bar{x})(x_j - \bar{x})) / (\Sigma_i(x_i - \bar{x})^2)$, where $w_{ij}$ are spatial weights and $W = \Sigma_i \Sigma_j w_{ij}$. Geary's C measures local dissimilarity: $C = ((n-1)/2W) \cdot (\Sigma_i \Sigma_j w_{ij}(x_i - x_j)^2) / (\Sigma_i(x_i - \bar{x})^2)$. Both metrics are computed using CuPy for dense and sparse matrices.

## Pathway and transcription factor activity inference

Rapids-singlecell provides GPU-accelerated implementations of enrichment methods from decoupler: multivariate linear model (mlm), univariate linear model (ulm), weighted aggregate (waggr), AUCell, and z-score. These implementations use CuPy for matrix operations and operate directly on GPU-resident data, avoiding host-device transfers when used within rapids-singlecell workflows.

## Batch correction with Harmony

Rapids-singlecell provides a GPU-accelerated implementation of the Harmony algorithm[21] for batch effect correction. The implementation operates on PCA embeddings and uses CuPy for all matrix operations, achieving several hundred-fold speedups over CPU baselines on million-cell datasets. Rapids-singlecells Harmony algorithm was completely re-engineered to maximize GPU throughput and memory efficiency. We redesigned the solver to operate directly on label vectors, thereby eliminating the computational and memory overhead of the dense one-hot encoding matrix φ. The implementation maintains a Pearson correlation of >95% for all corrected principal components compared to the reference baseline.

## Perturbation distance computation

Rapids-singlecell provides a GPU-accelerated implementation of the energy distance (E-distance) metric from pertpy through the *ptg* (pertpy-GPU) submodule. The *rapids_singlecell.ptg.Distance* class mirrors the pertpy *Distance* API and currently supports the E-distance metric, which quantifies transcriptomic differences between groups of cells in perturbation experiments such as CRISPR screens and drug treatments. The class operates on cell embeddings stored in *adata.obsm* (e.g., PCA coordinates) or expression matrices in *adata.layers*, and exposes *pairwise*, *onesided_distances*, and *bootstrap* methods compatible with pertpy's interface.

The energy distance between two groups X and Y is defined as:

$$E(X, Y) = 2\cdot\text{mean}(||X_i - Y_j||) - \text{mean}(||X_i - X_i'||) - \text{mean}(||Y_j - Y_j'||)$$

where the means are taken over all pairwise Euclidean distances between and within groups. This requires computing $O(n\cdot m)$ cross-group and $O(n^2 + m^2)$ within-group pairwise distances, making it a natural target for GPU parallelization.

The implementation uses a custom CUDA kernel to compute all required pairwise distance sums directly on the GPU. Cells are organized by group using a category-index mapping (category offsets and sorted cell index arrays), analogous to CSR-style indirection. For each group pair (i, j), the kernel computes the sum of Euclidean distances between all cells in group i and all cells in group j. The kernel employs tiled computation with configurable cell and feature tile sizes, shared memory for data reuse, and is JIT-compiled through CuPy. For within-group (diagonal) pairs, only the upper triangle of the distance matrix is accumulated, and groups with fewer than two cells are skipped. The raw distance sums are normalized by the appropriate pair counts ($n_i \cdot n_j$ for cross-group, $n_i \cdot (n_i - 1)/2$ for within-group) to obtain mean pairwise distances, from which the energy distance is assembled vectorially: $E[a,b] = 2\cdot d[a,b] - d[a,a] - d[b,b]$.

The E-distance implementation supports multi-GPU execution. Group pairs are split across available GPUs with load balancing based on group sizes, and each GPU computes its assigned subset of distance sums independently. Data (embeddings, category offsets, cell indices, and pair assignments) is replicated to each device using non-blocking CUDA streams for overlapped transfer and computation. After all kernels complete, partial sum matrices are aggregated on GPU 0 via cross-device copies.

For bootstrap confidence intervals, rather than precomputing an $n\times n$ cell distance matrix and sampling from it as in the CPU pertpy implementation, rapids-singlecell resamples cells and recomputes distances from scratch each iteration. Bootstrap indices are generated entirely on GPU using CuPy's random number generator: random floats are scaled by per-group sizes and converted to local indices, then mapped to global cell indices via group offsets - all without host-device transfers. This approach requires $O(n)$ rather than $O(n^2)$ memory and leverages multi-GPU parallelism for each bootstrap iteration. Bootstrap statistics (mean and variance) are computed from the stacked iteration results. The variance of the energy distance is approximated via the delta method: $\text{Var}[E(a,b)] \approx 4\cdot\text{Var}[d(a,b)] + \text{Var}[d(a,a)] + \text{Var}[d(b,b)]$.

The *onesided_distances* method computes distances from a single reference group (e.g., control) to all other groups by constructing only the necessary cross-group and diagonal pairs, reducing computation relative to a full pairwise matrix. Both pairwise and onesided results are returned as pandas DataFrames or Series with pertpy-compatible naming and indexing conventions.

## Custom GPU kernels

Performance-critical operations use hand-optimized CUDA RawKernels JIT-compiled through CuPy: scatter-add kernels with atomic operations for categorical aggregation;

centered Gram matrix construction with shared memory and float4 vectorized loads; pairwise distance kernels for squared Euclidean ($d_{ij} = ||x_i - x_j||^2$) and cosine distances ($d_{ij} = 1 - (x_i \cdot x_j)/(||x_i||\ ||x_j||)$); per-gene reduction kernels for mean and variance; and upper-triangle copy kernels for matrix symmetrization. All kernels operate directly on device memory, avoiding host-device transfers.

## Memory management and scaling

Memory allocation uses RMM with optional pooling and oversubscription. When datasets exceed GPU memory, RMM with enabled oversubscription spills to host memory with automatic paging. For datasets beyond single-GPU limits (~1 million cells with int32 sparse indices), Dask partitions AnnData objects and distributes computations across multiple GPUs or nodes. Generally, Dask partitions the expression matrix into chunks along the cell axis. Chunk sizes are chosen to balance memory constraints and computational efficiency: smaller chunks (e.g., 20,000 cells) reduce peak memory usage for out-of-core execution on memory-limited hardware, while larger chunks (e.g., 100,000 cells) improve throughput on high-memory multi-GPU systems by reducing scheduling overhead. For multi-GPU execution, the dataset can be rechunked so that partitions are distributed evenly across available devices. Optimal chunk sizes depend on available GPU memory, dataset dimensions, and the specific operations performed.

## Benchmarking

Benchmarks were designed to evaluate performance across varying data scales and hardware configurations using 10x Genomics datasets. We benchmarked the standard in-core analysis workflow (normalization, HVG selection, PCA, neighborhood graph construction, UMAP, and Leiden clustering) on a 1.3 million cell dataset. To evaluate batch correction scalability, we ran the Harmony algorithm on datasets of approximately 200,000, 2.4 million, and 11.4 million cells across single-GPU configurations. CPU baselines ran on an AMD Threadripper Pro 7975WX (32 cores, 64 threads) with 512GB RAM. Single-GPU experiments used NVIDIA DGX B200 (192 GB HBM3e), H200 (141 GB HBM3e), and NVIDIA RTX PRO 6000 Blackwell Workstation Edition (96 GB GDDR7) accelerators. Random seeds were fixed for all stochastic operations to ensure reproducibility across runs. Timing excluded I/O; each configuration was repeated three times and mean runtimes reported.

All benchmarks of rapids-singlecell (v0.13.5) were performed using scanpy (v1.12.0rc1) with anndata (v0.12.6) for data representation. GPU computations utilized CuPy (v14.0.0rc1) and the NVIDIA stack: cuML (v25.12.0), cuDF (v25.12.0), and RMM (v25.12.0). Clustering, neighbor search, dimensionality reduction, and batch correction employed leidenalg (v0.11.0), PyNNDescent (v0.5.13), UMAP (v0.5.9.post2), and harmonypy (v0.0.10), respectively. Numerical operations were backed by NumPy (v2.2.6) and SciPy (v1.16.3). We used scanpy (v1.12.0rc1) to maximise CPU performance. For CPU Leiden clustering we used the faster igraph backend which will become the default in an upcoming scanpy release, not the default leidenalg backend.

To demonstrate the feasibility of analyzing atlas-scale datasets, we benchmarked rapids-singlecell on the Tahoe 100M cell dataset. Using a node equipped with 8 NVIDIA

DGX B300 GPUs, the entire analysis pipeline - including log-normalization, highly variable gene (HVG) selection, scaling, PCA, nearest neighbor graph construction, UMAP, and Leiden clustering - was completed in under 20 minutes.

### Numerical validation

All implementations were validated against scanpy CPU baselines. Deterministic numerical operations (normalization, HVG selection, PCA) produce outputs within floating-point tolerance (numpy.allclose with default parameters). For stochastic algorithms (UMAP, t-SNE, Leiden clustering), we verified equivalent output distributions using adjusted Rand index for cluster assignments and preservation of local neighborhood structure for embeddings.

## Code and data availability

The rapids-singlecell source code is available at https://github.com/scverse/rapids_singlecell under the MIT license. Further documentation, tutorials and examples are available at https://rapids-singlecell.readthedocs.io/en/stable. Scripts, notebooks, and analysis results to reproduce our analysis, benchmarking, and figures are available at https://github.com/theislab/rapids_singlecell-reproducibility. Benchmarks used the 1.3 million mouse brain cell dataset from 10x Genomics (https://www.10xgenomics.com/datasets/1-3-million-brain-cells-from-e-18-mice-2-standard-1-3-0), a 11 million single-cell nuclei timelapse of mouse embryonic development dataset collection[22] (https://cellxgene.cziscience.com/collections/45d5d2c3-bc28-4814-aed6-0bb6f0e11c82), the human brain cell atlas 1.0 comprising 2.4 million cells[23] (https://cellxgene.cziscience.com/collections/283d65eb-dd53-496d-adb7-7570c7caa443), a 211 thousand cell cross tissue immune dataset[24] (https://cellxgene.cziscience.com/collections/62ef75e4-cbea-454e-a0ce-998ec40223d3), and the Tahoe-100M dataset (https://huggingface.co/datasets/tahoebio/Tahoe-100M)[18].

## Acknowledgments

The authors thank all users of rapids-singlecell who have provided valuable feedback. The authors acknowledge technical support from NVIDIA and scverse. We would also like to thank Seulah Park for creating the rapids-singlecell logo.

## Author contributions

SD conceived the study. SD implemented rapids-singlecell with help from LH, PA, IG, IV, FF, and CN. SD, LH, LM, and SJ wrote the manuscript and designed the figures together with FJT. MG assisted with benchmarking. TJC, MB, FJT supervised the work. All authors read, corrected and approved the final manuscript.

## Competing interests

SD, MG, TJ & CN are employees of NVIDIA, which develops hardware and software relevant to this work. LH is an employee of LaminLabs. FJT consults for Immunai Inc., Singularity Bio B.V., CytoReason Ltd, and Omniscope Ltd, and has ownership interest in Dermagnostix GmbH and Cellarity.